\documentclass{article}
\usepackage{amsmath}
\usepackage{newtxtext, newtxmath}
\usepackage{graphicx}
\usepackage{geometry}
\usepackage{hyperref}
\usepackage{indentfirst}
\usepackage[numbers,square,sort&compress]{natbib}
\usepackage{siunitx}

\title{Vector resolved energy fluxes and collisional energy losses \\ in magnetic nozzle radiofrequency plasma thrusters}
\author{Kazuma Emoto,\textsuperscript{1,\thanks{kazuma-emoto-vh@ynu.jp}\ } 
        Kazunori Takahashi,\textsuperscript{2} 
        and Yoshinori Takao\textsuperscript{3,\thanks{takao@ynu.ac.jp}}}
\date{}

\begin{document}

\maketitle

\begin{center}
    \textsuperscript{1}\textit{Department of Mechanical Engineering, Materials Science, and Ocean Engineering, Yokohama National University, Yokohama 240-8501, Japan}
    
    \textsuperscript{2}\textit{Department of Electrical Engineering, Tohoku University, Sendai 980-8579, Japan}
    
    \textsuperscript{3}\textit{Division of Systems Research, Yokohama National University, Yokohama 240-8501, Japan}
\end{center}

\vspace{8pt}

\begin{abstract}
    Energy losses in a magnetic nozzle radiofrequency plasma thruster are investigated to improve the thruster efficiency, which are calculated from particle energy losses in fully kinetic simulations. The simulations calculate particle energy fluxes with a vector resolution including the plasma energy lost to the dielectric wall, the plasma beam energy, and the divergent plasma energy in addition to collisional energy losses. As a result, distributions of energy losses in the thruster and the ratios of the energy losses to the input power are obtained. The simulation results show that the plasma energy lost to the dielectric is dramatically suppressed by increasing the magnetic field strength and the ion beam energy increases instead. In addition, the divergent ion energy and collisional energy losses account for approximately 4--12\% and 30--40\%, respectively, regardless of the magnetic field strength. 
\end{abstract}

\vspace{12pt}



\section*{INTRODUCTION}

Magnetic nozzle radiofrequency (rf) plasma thrusters have been developed worldwide for future high-power electric propulsion systems \cite{Charles2009_jpd, Merino2016_psst, Little2019_prl, Takahashi2019_rmpp, Chen2020_pre, Doyle2020_fp, Wachs2020_psst}. The main components of the magnetic nozzle rf plasma thruster are an rf antenna, a dielectric tube, and the solenoid. The thruster does not have a cathode or an external neutralizer, i.e., a completely electrodeless configuration. Therefore, electrode wear which limits the lifetime does not occur, enabling the long lifetime in space. The magnetic nozzle rf plasma thrusters are expected to be utilized instead of conventional electric thrusters, such as ion and Hall thrusters, as the long-term operational propulsion system.

In the magnetic nozzle rf plasma thrusters, the electric power is delivered to the plasma from the rf antenna wound around the source tube, where the solenoid or the permanent magnets are set to produce the magnetic field in the source and the magnetic nozzle \cite{Chen2007_ppcf, Takahashi2008_pop, Charles2009_jpd, Virko2010_psst}. The plasma is accelerated through the magnetic nozzle toward the downstream direction and obtains the axial momentum. In the magnetic nozzle, an azimuthal current is induced by the diamagnetic effect and produces the axial Lorentz force with the radial magnetic field \cite{Ahedo2010_pop, Fruchtman2012_pop, Takahashi2013_prl, Takahashi2016_psst, Emoto2021_pop}.

Thruster efficiencies of magnetic nozzle rf plasma thrsuters are summarized in \cite{Takahashi2019_rmpp}, and a recent study reported the thruster efficiency approaching 20\% at maximum for the rf power of 6 kW \cite{Takahashi2021_sr}. However, Hall thrusters show the thruster efficiency of 35--60\% \cite{Goebel2008_fundamentals}, which have been already utilized successfully on satellites and space probes \cite{Cara2005_aa, Pidgeon2006_aiaa}. The thruster efficiency of magnetic nozzle rf plasma thrusters is still relatively poor compared to that of Hall thrusters, although magnetic nozzle rf plasma thrusters have the advantage of not requiring cathodes and an external neutralizer. Therefore, further improvement of the thruster efficiency is an important challenge.

Considering that Hall thrusters achieve the thruster efficiency of 60\% at maximum, magnetic nozzle rf plasma thrusters do not utilize the input power efficiently compared with Hall thrusters.  In magnetic nozzle rf plasma thrusters, the power is supplied from only the rf antenna because of the completely electrodeless configuration. The rf power is mainly absorbed by electrons in the inductively coupled or helicon modes. Note that the neutral energy is negligibly small because the neutral temperature is roughly 300 K, while the measured ion and electron temperatures are 0.1--1 eV \cite{Scime2000_pop} and 1--20 eV \cite{Takahashi2011_prl, Aguirre2020_pop}, respectively. In addition, the magnetostatic field produced by the solenoid supplies no energy to the plasma. Therefore, to impart the axial momentum to ions and obtain the thrust, the electron energy heated by the rf electromagnetic fields should be converted to the axial ion energy in the magnetic nozzle. In the case of the low thruster efficiency, the electron energy does not converted to the axial ion energy efficiently and is expected to be lost in the thruster. Here, possible energy losses are energy fluxes on the dielectric wall, a divergent plasma energy, a rotation plasma energy, and collisional energy losses. To improve the thruster efficiency of magnetic nozzle rf plasma thrusters, it is necessary to know where and how the energies are lost in the thruster and to identify the major process of the energy loss. 

The energy balance in the thruster has been analyzed by a global discharge model \cite{Lafleur2014_pop}, implying the significant energy losss to the wall. An individual measurement of the axial force imparted to the lateral dielectric wall showed that the non-negligible axial momentum is lost to there in magnetic nozzle rf plasma thrusters \cite{Takahashi2015_prl}. Moreover, the energy lost to the lateral dielectric wall was measured by using a momentum vector measurement instrument and Langmuir probes, which accounted for 20--60\% of the input power depending on the magnetic field strength \cite{Takahashi2020_sr}. However, other energy losses except the energy lost to the dielectric wall have not been fully understood yet.

In this study, fully kinetic simulations of a magnetic nozzle rf plasma thruster are conducted to investigate the energy losses in the thruster. The simulations obtain energy fluxes by ions and electrons with a vector resolution, which are calculated directly from particle motions in the simulations. Collisional energy losses are also calculated in the simulations. Firstly, the calculation model of fully kinetic simulations employed in this paper is briefly described. Then, the energy losses in the thruster and these ratios to the input power are reported. The simulation results show that the ion and electron energies lost to the dielectric wall are suppressed by increasing the magnetic field strength and the thruster efficiency dramatically increases instead, which are consistent with previous experiments \cite{Takahashi2020_sr}.

\section*{NUMERICAL MODEL}

We have employed two-dimensional ($x$-$y$) and bidirectional calculation model to reduce the calculation cost \cite{Emoto2021_pop, Emoto2021_psst}. Although bidirectional thrusters are different from the common thruster, they are proposed for the space debris removal \cite{Takahashi2018_sr}. The numerical model has an infinite length in the $z$-direction, and the simulation results are discussed in the unit length of the $z$-direction. The plasma kinetics in the thruster is analyzed using particle-in-cell and Monte calro collisions (PIC-MCC) techniques. Details of the calculation model and simulation method were given in previous papers \cite{Takao2015_pop, Takase2018_pop, Emoto2021_pop, Emoto2021_psst}, and a brief description is written in this paper.

\begin{figure}
    \centering
    \includegraphics{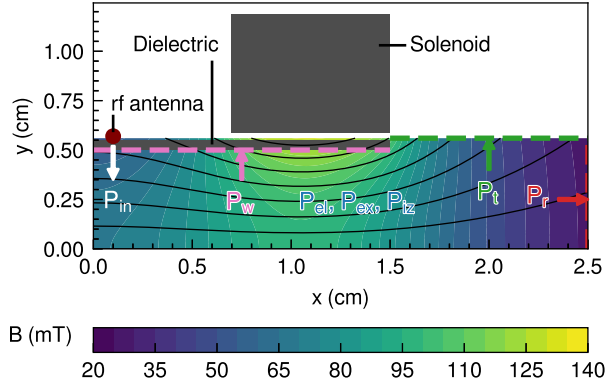}
    \caption{A schematic of the calculation area and a magnetic field strength produced by the solenoid current of 2.0 kA. Solid black lines show the magnetic field lines. Energy fluxes and collisional energy losses in the magnetic nozzle rf plasma thruster are also plotted. $P_{in}$ is the input power from the rf antenna (a white arrow), $P_{w}$, $P_{r}$, and $P_{t}$ are energy losses per unit time by colliding with the lateral dielectric wall (a pink arrow), by colliding with the right boundary at $x$ = 2.5 cm (a red arrow), by colliding with the top boundary at $y$ = 0.56 cm (a green arrow), respectively. These energy losses per unit time are calculated from particle motions for ions and electrons and $x$-, $y$-, and $z$-directions independently. $P_{el}$, $P_{ex}$, and $P_{iz}$ are the elastic, excitation, and ionization energy losses per unit time, respectively (a blue font). Because left and bottom boundaries are assumed to be symmetry, particles are reflected on the boundaries, and the particle energies are not lost.}
    \label{fig:solenoid_magnetic_field}
\end{figure}

Figure \ref{fig:solenoid_magnetic_field} shows a schematic of the calculation area employed in this paper. The calculation area is 2.5 cm $\times$ 0.56 cm including the dielectric and divided into 50 \si{\micro m} $\times$ 50 \si{\micro m} cells. The solenoid produces the magnetic field lines as shown by solid black lines in Figure \ref{fig:solenoid_magnetic_field}. The solenoid current is set to 0.1, 0.4, and 2.0 kA to investigate the dependence of the magnetic field strength. The magnetic field strength at 2.0 kA is also shown as a colormap in Figure \ref{fig:solenoid_magnetic_field}. The plasma absorbs the rf power supplied from the rf antenna and maintains the discharge. Ions and electrons are lost to the lateral dielectric wall, the right boundary at $x$ = 2.5 cm, or the top boundary at $y$ = 0.56 cm. Because the calculation area is symmetry about the $x$- and $y$-axes, ions and electrons are reflected on the left boundary at $x$ = 0 cm and the bottom boundary at $y$ = 0 cm. The motions of ions and electrons are solved by using the Boris method \cite{Birdsall2004_crc}. In the simulations, singly charged xenon ions Xe$^+$ and electrons e$^-$ are treated. Their time steps are set to 0.125 ns and 3.57 ps, respectively. In addition, electron-neutral collisions including elastic, excitation, and ionization collisions are solved by using null-collision method \cite{Vahedi1995_cpc}. The neutral density is spatiotemporary constant in the simulations and set to \SI{2.0e19}{m^{-3}}. The rf frequency is set to 80 MHz, and the power absorption is controlled to be 3.5 W, where the unit length is assumed in the $z$-direction. Electric and magnetic fields employed in this paper are the electrostatic field $E_{es}$ generated by charged particles and the surface charge on the dielectric, the electromagnetic field $E_{em}$ induced by the rf antenna, and the magnetostatic field $B$ produced by the solenoid. These fields are solved by using fast Fourier transformation. The boundary conditions of the electrostatic field are Dirichlet boundary conditions at the right and top boundaries ($E_{es}$ = 0) and the Neumann boundary conditions at the left and bottom boundaries ($\partial E_{es}/\partial n$ = 0 with $\partial/\partial n$ being the normal derivative). Note that the Dirichlet boundary conditions generate the sheath near the right and top boundaries, as shown in our previous paper \cite{Emoto2021_psst}. Simulation results are averaged over 30 \si{\micro s}.

Here, the magnetic field strengths under the solenoid are 5, 20, and 100 mT for the solenoid currents of 0.1, 0.4, and 2.0 kA, respectively. For the magnetic field strength of 100 mT, the ion Larmor radius $r_L$ becomes approximately 1.3 cm, where the ion temperature is assumed to be 0.5 eV. Then, the ratio $r_L/L$ is 1.3, where $L$ is the thruster height of 1 cm  (diameter in cylindrical coordinates). Therefore, ions are not fully magnetized for all solenoid currents. 

Figure \ref{fig:solenoid_magnetic_field} also shows energy fluxes and collisional energy losses considered in the simulations. In this study, energy losses are evaluated by using those per unit time (i.e., the power). $P_{in}$ is the input energy from the rf antenna, $P_{w}$ is the energy loss by colliding with the lateral dielectric wall, $P_{r}$ is the energy loss by colliding with the right boundary at $x$ = 2.5 cm, and $P_{t}$ is the energy loss by colliding with the top boundary at $y$ = 0.56 cm. These energy losses are calculated from particle motions for ions and electrons and $x$-, $y$-, and $z$-directions  independently. $P_{el}$, $P_{ex}$, and $P_{iz}$ are the elastic, excitation, and ionization energy losses, respectively. Because left and bottom boundaries are assumed to be symmetry, particles are reflected on the boundaries, and the particle energies are not lost.

The energy losses considered in this study are summarized in Table \ref{tab:energy_definition}, and the subscripts are described in Table \ref{tab:energy_subscript}. The energy flux of the particle species $\alpha$ to the boundary $\beta$ in the $\gamma$-direction $p_{\alpha,\beta,\gamma}$ is calculated by
\begin{equation}
    p_{\alpha,\beta,\gamma} = \frac{\sum_i (m_\alpha v_{\alpha,\beta,\gamma,i}^2 / 2)}{\Delta S \Delta t_a},
\end{equation}
where $i$ is the particle index, $m_\alpha$ is the mass of the particle species $\alpha$, $v_{\alpha,\beta,\gamma,i}$ is the velocity of the particle $i$ in the $\gamma$-direction, $\Delta S$ is the cell area, and $\Delta t_a$ is the averaging time. Here, $\alpha$ is $i$ or $e$, $\beta$ is $r$, $t$, or $w$, and $\gamma$ is $x$, $y$, or $z$, according to Table \ref{tab:energy_subscript}. It should be noted the sheaths at the boundaries are self-consistently solved in the simulation; hence the energy fluxes calculated in the simulations contain the sheath effect. The collisional energy loss density $p_{c}$ is written as
\begin{equation}
    p_{c} = \frac{\sum \Delta E}{\Delta V \Delta t_a},
\end{equation}
where $\Delta E$ is the energy loss by a collision and $\Delta V$ is the cell volume. Here, the subscript $c$ is $el$, $ex$, or $iz$ according to Table \ref{tab:energy_subscript}. The total energy losses per unit time $P_{\alpha,\beta,\gamma}$ and $P_{c}$ are calculated by integrating $p_{\alpha,\beta,\gamma}$ and $p_{c}$ over the area and volume, respectively, i.e., given by
\begin{equation}
    \label{eq:sum_surface_energy_loss}
    P_{\alpha,\beta,\gamma} = \sum p_{\alpha,\beta,\gamma} \Delta S,
\end{equation}
\begin{equation}
    \label{eq:sum_volume_energy_loss}
    P_{c} = \sum p_{c} \Delta V.
\end{equation}
The sum $P_{\alpha,\beta}$ of the energy losses per unit time in all directions is written as
\begin{equation}
    P_{\alpha,\beta} = P_{\alpha,\beta,x} + P_{\alpha,\beta,y} + P_{\alpha,\beta,z}.
\end{equation}
Note that the energy fluxes and collisional energy loss density are dependent on the size of the calculation area, indicating that this paper evaluates the thruster performance in a small vacuum chamber of 2.5 cm $\times$ 0.56 cm. 

\begin{table}
    \centering
    \caption{Definitions of energy losses considered in this study.}
    \label{tab:energy_definition}
    \begin{tabular}{ll}
        \hline
        Symbols & Descriptions \\
        \hline
        $p_{i,r,x}$ & Ion energy loss in the $x$-direction by colliding with the right boundary at $x$ = 2.5 cm \\
        $p_{i,r,y}$ & Ion energy loss in the $y$-direction by colliding with the right boundary at $x$ = 2.5 cm \\
        $p_{i,r,z}$ & Ion energy loss in the $z$-direction by colliding with the right boundary at $x$ = 2.5 cm \\
        $p_{e,r,x}$ & Electron energy loss in the $x$-direction by colliding with the right boundary at $x$ = 2.5 cm \\
        $p_{e,r,y}$ & Electron energy loss in the $y$-direction by colliding with the right boundary at $x$ = 2.5 cm \\
        $p_{e,r,z}$ & Electron energy loss in the $z$-direction by colliding with the right boundary at $x$ = 2.5 cm \\
        $p_{i,t,x}$ & Ion energy loss in the $x$--direction by colliding with the top boundary at $y$ = 0.56 cm \\
        $p_{i,t,y}$ & Ion energy loss in the $y$-direction by colliding with the top boundary at $y$ = 0.56 cm \\
        $p_{i,t,z}$ & Ion energy loss in the $z$-direction by colliding with the top boundary at $y$ = 0.56 cm \\
        $p_{e,t,x}$ & Electron energy loss in the $x$--direction by colliding with the top boundary at $y$ = 0.56 cm \\
        $p_{e,t,y}$ & Electron energy loss in the $y$-direction by colliding with the top boundary at $y$ = 0.56 cm \\
        $p_{e,t,z}$ & Electron energy loss in the $z$-direction by colliding with the top boundary at $y$ = 0.56 cm \\
        $p_{i,w}$ & Ion energy loss by colliding with the dielectric wall at $y$ = 0.5 cm \\
        $p_{e,w}$ & Electron energy loss by colliding with the dielectric wall at $y$ = 0.5 cm  \\
        $p_{el}$ & Elastic energy loss \\
        $p_{ex}$ & Excitation energy loss \\
        $p_{iz}$ & Ionization energy loss \\
        \hline
    \end{tabular}
\end{table}

\begin{table}
    \centering
    \caption{Subscripts of energy losses.}
    \label{tab:energy_subscript}
    \begin{tabular}{ll}
        \hline
        Symbols & Descriptions \\
        \hline
        $i$ & ion \\
        $e$ & electron \\
        $r$ & right boundary \\
        $t$ & top boundary \\
        $w$ & dielectric wall \\
        $x$ & $x$-direction \\
        $y$ & $y$-direction \\
        $z$ & $z$-direction \\
        $el$ & elastic \\
        $ex$ & excitation \\
        $iz$ & ionization \\
        \hline
    \end{tabular}
\end{table}

The plasma power contributing to the thrust $P_{th}$ is defined as

\begin{equation}
    \label{eq:thrust_energy}
    P_{th} = P_{i,r,x} + P_{e,r,x} + P_{i,t,x} + P_{e,t,x}.
\end{equation}

\noindent
Then, the thruster efficiency $\eta$ in the simulations is calculated as

\begin{equation}
    \label{eq:thruster_efficiency}
    \eta = \frac{P_{th}}{P_{in}}.
\end{equation}

\section*{RESULTS AND DISCUSSION}

\begin{figure}
    \centering
    \includegraphics{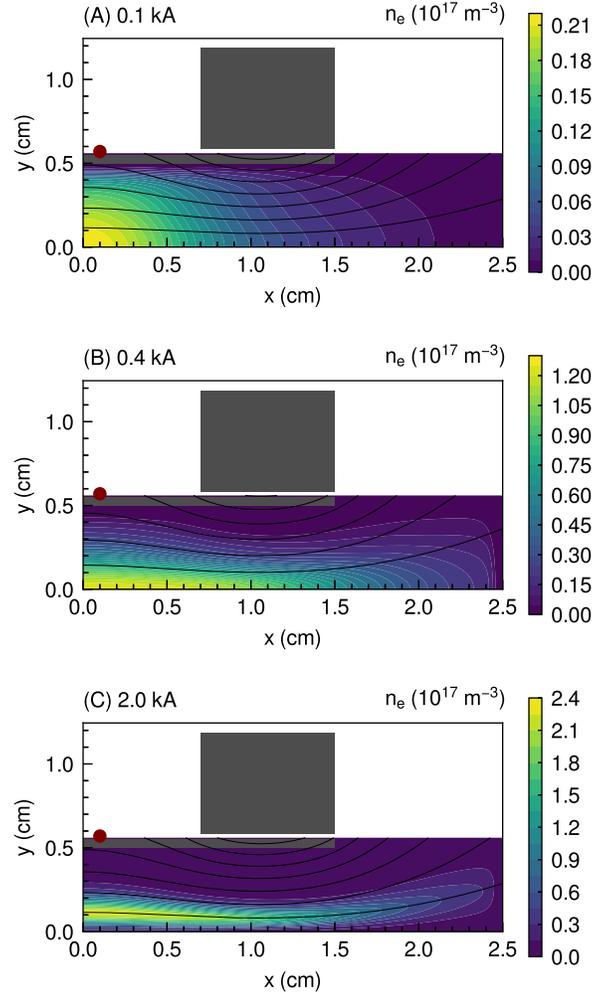}
    \caption{$x$-$y$ profiles of the electron number density $n_e$ for three solenoid currents of (A) 0.1, (B) 0.4, and (C) 2.0 kA. Solid black lines show magnetic field lines.}
    \label{fig:electron_number_density}
\end{figure}

Figure \ref{fig:electron_number_density} shows $x$-$y$ profiles of the electron number density $n_e$ for the three solenoid currents of 0.1, 0.4, and 2.0 kA. The electron number density $n_e$ is large near the axial location of the rf antenna ($x$ = 0.1 cm) and decreases in the downstream direction due to the magnetic expansion, the plasma diffusion, the ion acceleration, and the electron reflection by the electric field. The plasma distribution is confined to the center of the magnetic nozzle for the 0.4 kA turn case in Figure \ref{fig:electron_number_density}(B), and the off-axis density peak can be seen for the 2.0 kA turn case as in Figure \ref{fig:electron_number_density}(C). These density-profile transitions by increasing the magnetic field strength were measured in previous experiments \cite{Takahashi2009_apl, Charles2010_apl, Ghosh2017_pop, Gulbrandsen2017_fp, Takahashi2017_pop}, indicating that the simulation results reproduce the experimental one qualitatively. 

\begin{figure}
    \centering
    \includegraphics{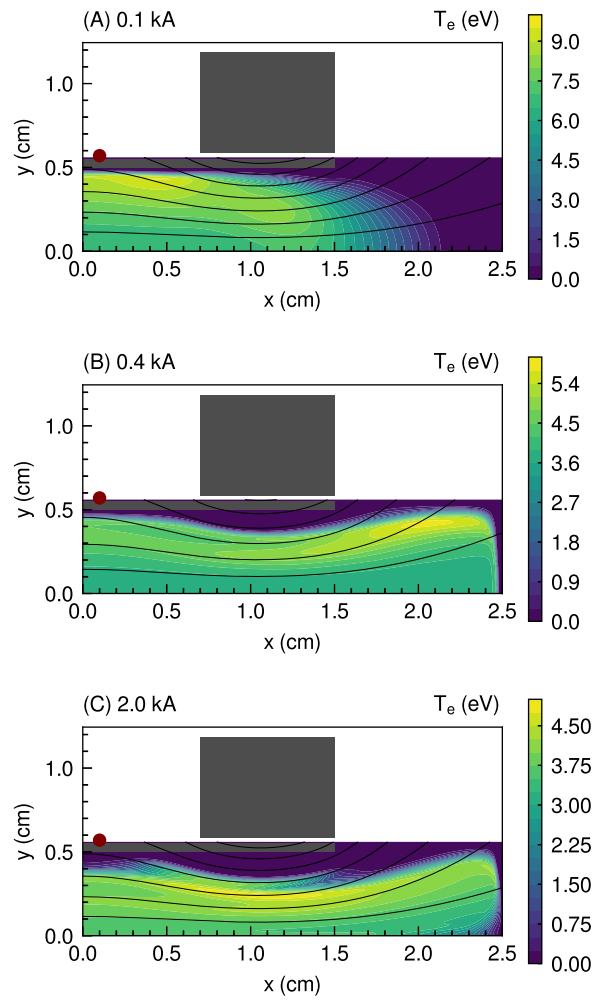}
    \caption{$x$-$y$ profiles of the electron temperature $T_e$ for the three solenoid currents of (A) 0.1, (B) 0.4, and (C) 2.0 kA. Solid black lines show magnetic field lines.}
    \label{fig:electron_temperature}
\end{figure}

Figure \ref{fig:electron_temperature} shows $x$-$y$ profiles of the electron temperature $T_e$ for the three solenoid currents of 0.1, 0.4, and 2.0 kA. As shown in Figure \ref{fig:electron_temperature}, the electron temperature is high on the magnetic field lines passing through the rf antenna for the three solenoid currents cases. For the 0.1 kA case, the electron temperature decreases in the downstream region within 1.5 cm < $x$ < 2.5 cm. Note that the electrons for the 0.1 kA case are less magnetized because the Larmor radius at the center of the solenoid is approximately 0.17 cm using the magnetic field strength of 5 mT and the electron temperature of 5 eV, less transported along the magnetic field lines. For the 0.4 and 2.0 kA cases, however, the high-temperature electrons are transported to the downstream region through the magnetic nozzle and expected to be lost to the downstream boundaries.

\begin{figure}
    \centering
    \includegraphics{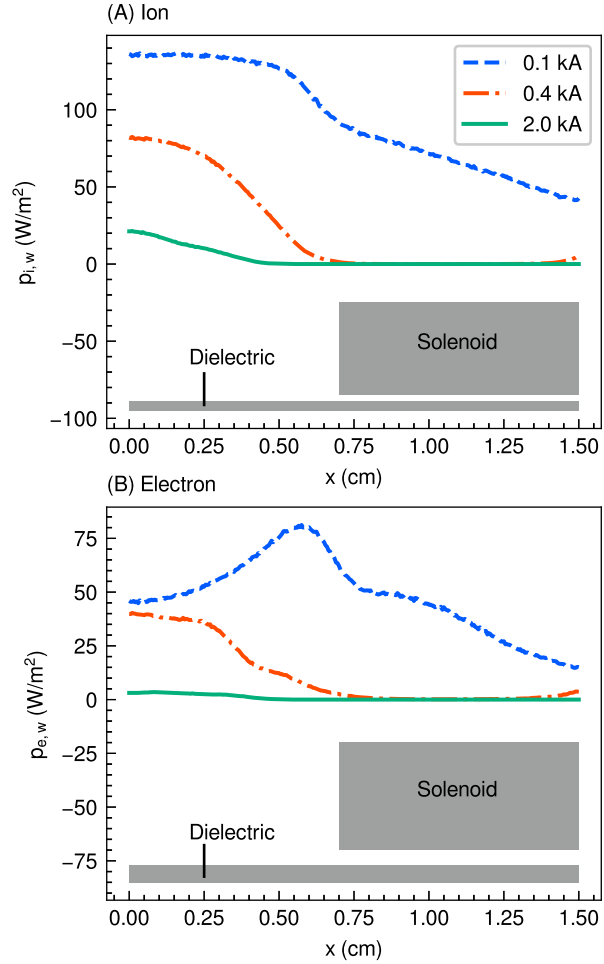}
    \caption{$x$ profile of the (A) ion and (B) electron energy flux on the lateral dielectric wall $p_{i,w}$ and $p_{e,w}$, respectively for three solenoid currents of 0.1 (a dashed blue line), 0.4 (a dotted-dashed orange line), and 2.0 kA (a solid green line). Grey boxes shows locations of the solenoid and the lateral dielectric wall.}
    \label{fig:energy_flux_tube}
\end{figure}

Figure \ref{fig:energy_flux_tube} shows $x$ profiles of the ion and electron energy fluxes on the lateral dielectric wall $p_{i,w}$ and $p_{e,w}$, respectively, for three solenoid currents of 0.1, 0.4, and 2.0 kA. The ion and electron energy fluxes on the lateral dielectric wall are large at $x$ = 0--0.7 cm and decrease with increasing the solenoid current. The large electron energy flux at $x$ = 0--0.7 cm for the 0.1 kA turn case results from the fact that the electrons are somewhat magnetized and tend to move along the magnetic field lines, which intersect the lateral dielectric wall as shown in Figure \ref{fig:electron_number_density}. In addition, the decrease of the electron energy flux is due to the confinement by the strong magnetic field. However, the ions are not fully magnetized in this simulation and do not move only along the magnetic field lines. It is expected that the ion motions are affected by the potential distribution as previously observed in experiments \cite{Cox2008_apl, Takahashi2009_apl_2}. 

\begin{figure}
    \centering
    \includegraphics{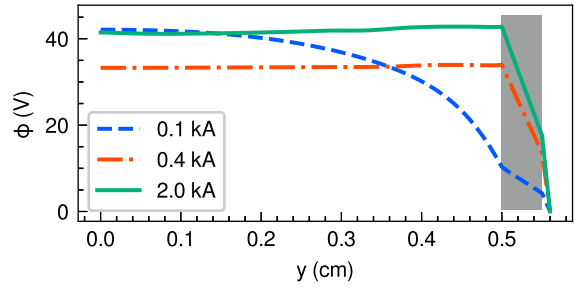}
    \caption{$y$ profile of the potential $\phi$ at $x$ = 1.1 cm for three solenoid currents of 0.1 (a dashed blue line), 0.4 (a dotted-dashed orange line), and 2.0 kA (a solid green line). A grey box shows the location of the dielectric wall.}
    \label{fig:scalar_potential}
\end{figure}

\begin{figure}
    \centering
    \includegraphics{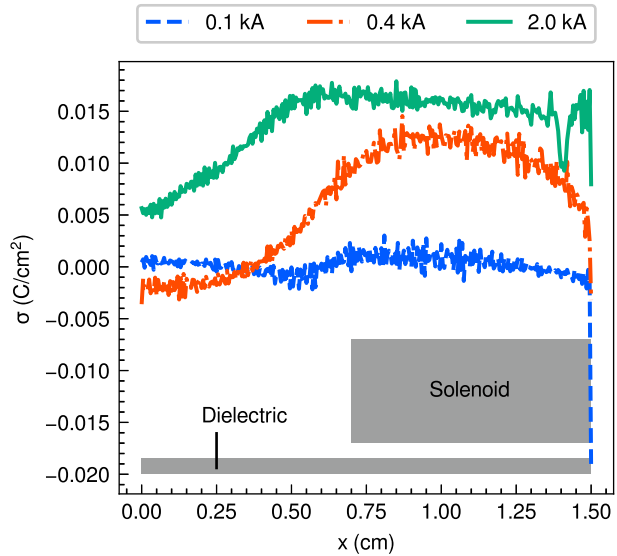}
    \caption{$x$ profile of the surface charge density on the lateral dielectric wall $\sigma$ for three solenoid currents of 0.1 (a dashed blue line), 0.4 (a dotted-dashed orange line), and 2.0 kA (a solid green line). Grey boxes shows locations of the solenoid and the lateral dielectric wall.}
    \label{fig:surface_charge_density}
\end{figure}

Figure \ref{fig:scalar_potential} shows $y$ profile of the potential $\phi$ at $x$ = 1.1 cm for the three solenoid currents of 0.1, 0.4, and 2.0 kA. The radial potential drop is clearly formed around $y$ = 0.2--0.5 cm for the 0.1 kA case; the ions are accelerated in the radial direction there and would take the significant energy from the system to the wall. For the 0.4 and 2.0 kA cases, the slight increases in the potential along the $y$ axis can be seen, and the wall potentials are higher than that in the plasma core, suppressing the ion loss to the dielectric. These structures can be interpreted as a result of the inhibition of the electron transport to the wall by the magnetic field, resulting in the positive charge up of the wall by the ions. To have a confidence on the positive charge up, $x$ profile of the surface charge $\sigma$ on the wall is analyzed and shown in Figure \ref{fig:surface_charge_density}, demonstrating the positively charged up wall at the solenoid location, where the magnetic field lines are almost parallel to the wall.

\begin{figure}
    \centering
    \includegraphics{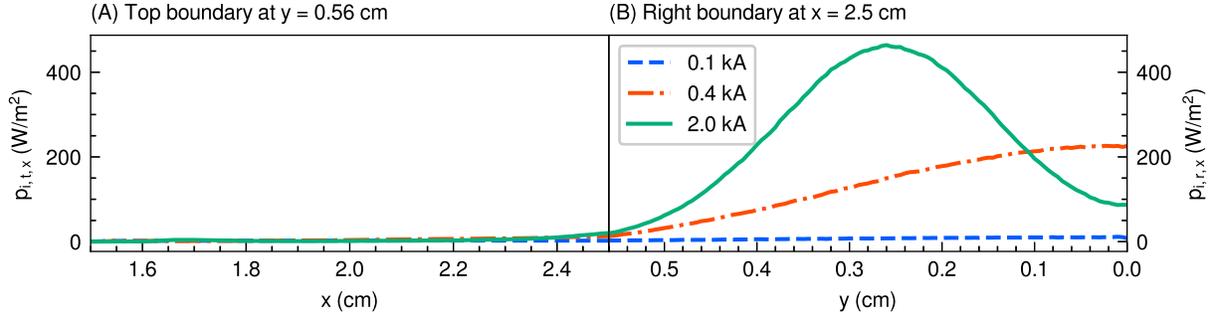}
    \caption{(A) $x$ and (B) $y$ profiles of the ion energy fluxes in the $x$-direction on the top boundary ($y$ = 0.56 cm) $p_{i,t,x}$ and the right boundary ($x$ = 2.5 cm) $p_{i,r,x}$ for three solenoid currents of 0.1 (a dashed blue line), 0.4 (a dotted-dashed orange line), and 2.0 kA (a solid green line). Note that $x$ = 2.5 cm in (A) corresponds to $y$ = 0.56 cm in (B)}
    \label{fig:ion_energy_flux_north_and_east_1}
\end{figure}

Figure \ref{fig:ion_energy_flux_north_and_east_1} shows $x$ and $y$ profiles of the ion energy fluxes in the $x$-direction on the top boundary $p_{i,t,x}$ and the right boundary $p_{i,r,x}$, respectively, for the three solenoid currents of 0.1, 0.4, and 2.0 kA. Note that the ion energy flux $p_{i,t,x}$ and $p_{i,r,x}$ contribute to the increase in the thrust. As shown in Figure \ref{fig:ion_energy_flux_north_and_east_1}(A), the ion energy flux in the $x$-direction on the top boundary is negligibly small. However, the ion energy flux in the $x$-direction on the right boundary $p_{i,r,x}$ significantly increases with increasing the solenoid current and achieves 464 \si{W/m^2} at maximum as shown in Figure \ref{fig:ion_energy_flux_north_and_east_1}(B), while it decreases at $y$ = 0--0.1 cm when increasing the solenoid current from 0.4 to 2.0 kA. The change in the position of the maximum ion energy flux and the low energy flux around $y$ = 0 cm in Figure \ref{fig:ion_energy_flux_north_and_east_1}(B) are consistent with the distributions of the electron number density $n_e$ in Figure \ref{fig:electron_number_density}. 

\begin{figure}
    \centering
    \includegraphics{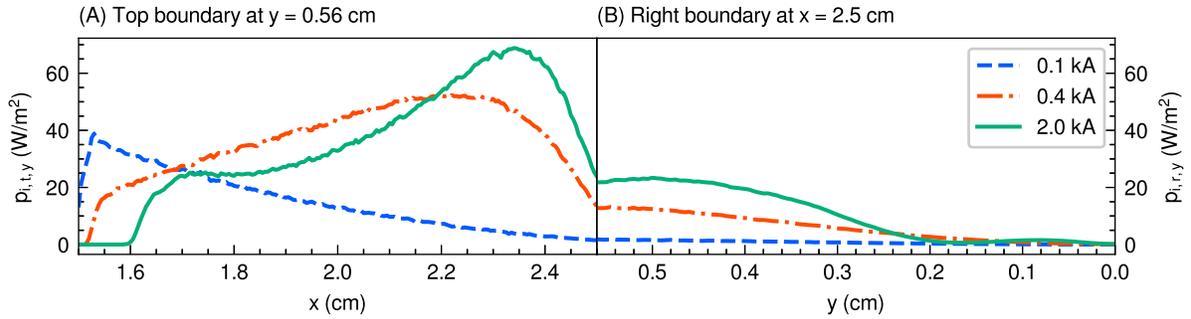}
    \caption{(A) $x$ and (B) $y$ profiles of the ion energy fluxes in the $y$-direction on the top boundary ($y$ = 0.56 cm) $p_{i,t,y}$ and the right boundary ($x$ = 2.5 cm) $p_{i,r,y}$, respectively, for three solenoid currents of 0.1 (a dashed blue line), 0.4 (a dotted-dashed orange line), and 2.0 kA (a solid green line). Note that $x$ = 2.5 cm in (A) corresponds to $y$ = 0.56 cm in (B)}
    \label{fig:ion_energy_flux_north_and_east_2}
\end{figure}

Figure \ref{fig:ion_energy_flux_north_and_east_2} shows $x$ and $y$ profiles of the ion energy fluxes in the $y$-direction on the top boundary $p_{i,t,y}$ and the right boundary $p_{i,r,y}$, respectively, for the three solenoid currents of 0.1, 0.4, and 2.0 kA. The ion energy fluxes in the $y$-direction $p_{i,t,y}$ and $p_{i,r,y}$ are 0--69 W/m$^2$. By changing the solenoid current from 0.1 to 0.4 kA, the position of the maximum ion energy flux in the $y$-direction shifts to the positive $x$-direction in Figure \ref{fig:ion_energy_flux_north_and_east_2}(A), and the maximum energy loss increases. The shift and increase of the maximum energy losses are because the plasma is localized near the center region around $x$ axis for the larger solenoid current as shown in Figure \ref{fig:electron_number_density} and are transported along the magnetic field lines.

\begin{figure}
    \centering
    \includegraphics{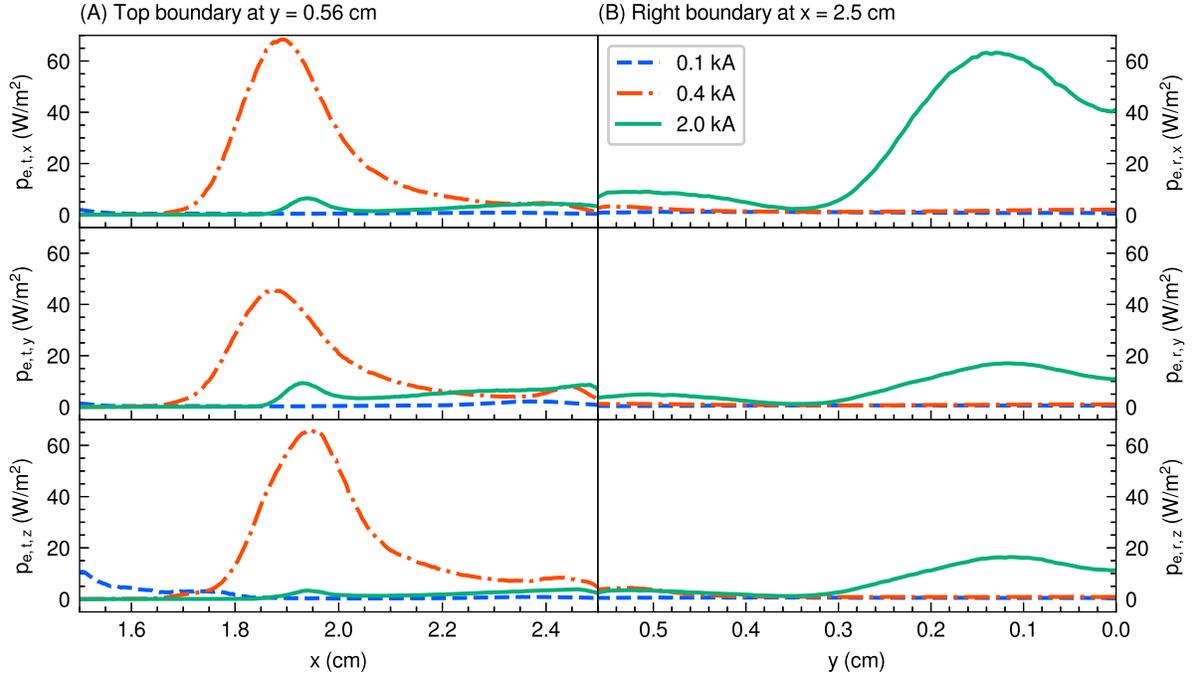}
    \caption{(A) $x$ profile of the electron energy fluxes on the top boundary ($y$ = 0.56 cm) $p_{e,t,x}$, $p_{e,t,y}$, and $p_{e,t,z}$ and (B) $y$ profile of the electron energy flux on the right boundary ($x$ = 2.5 cm) $p_{e,r,x}$, $p_{e,r,y}$, and $p_{e,r,z}$, for the three solenoid currents of 0.1 (a dashed blue line), 0.4 (a dotted-dashed orange line), and 2.0 kA (a solid green line). Note that $x$ = 2.5 cm in (A) corresponds to $y$ = 0.56 cm in (B)}
    \label{fig:electron_energy_flux_north_and_east}
\end{figure}

Figure \ref{fig:electron_energy_flux_north_and_east} shows $x$ profile of the electron energy fluxes on the top boundary ($p_{e,t,x}$, $p_{e,t,y}$, and $p_{e,t,z}$) and $y$ profile of the electron energy fluxes on the right boundary ($p_{e,r,x}$, $p_{e,r,y}$, and $p_{e,r,z}$) for the three solenoid currents of 0.1, 0.4, and 2.0 kA. Note that the electron energy fluxes $p_{e,t,x}$ and $p_{e,r,x}$ contribute to the increase in the thrust. The electron energy fluxes for the 0.1 kA case are sufficiently small compared with the 0.4 and 2.0 kA cases, indicating that the electron energy fluxes on the top and right boundaries increase with increasing the solenoid current. This increase in the energy fluxes is because the magnetic nozzle enhances the electron transport to the downstream region and suppresses the energy losses to the dielectric as shown in Figures \ref{fig:energy_flux_tube}. For the 0.4 kA case, the location of the maximum electron energy flux is $x$ = 1.9 cm in Figure \ref{fig:electron_energy_flux_north_and_east}(A), because the energetic electrons are magnetized and transported along the magnetic field lines passing through the location under the rf antenna, as reported in \cite{Takahashi2009_apl, Zhang2016_pop, Takahashi2017_pop, Bennet2019_pop}, where the electrons are considerably heated by the rf field. As shown in Figure \ref{fig:electron_temperature}, the high-temperature electrons for the 0.4 kA case pass through the top boundary at $y$ = 0.56 cm along the magnetic field lines, corresponding to the electron energy flux shown in Figure \ref{fig:electron_energy_flux_north_and_east}(A). 

The electron energy fluxes to the right boundary for the 2.0 kA case are much larger than those for the 0.1 and 0.4 cases as seen in Figure \ref{fig:electron_energy_flux_north_and_east}(B), whereas the significant energy fluxes to the top boundary appear for the 0.4 kA case as seen in Figure \ref{fig:electron_energy_flux_north_and_east}(A). These energy fluxes would be dominated by the density and temperature profiles in Figures \ref{fig:electron_number_density} and \ref{fig:electron_temperature}, respectively. The simulation results in Figure \ref{fig:electron_energy_flux_north_and_east} show that the electron energy is well transported to the right boundary in the inner region along the magnetic nozzle for the strong magnetic field case of 2.0 kA. For the 0.4 kA case, however, the significant electron energy is transported to the top boundary along the divergent magnetic field lines. The detailed particle and energy transport processes have not been fully understood yet and remain further investigation.

In the simulations, the ion and electron energy fluxes in the $x$-direction are considered to be the plasma beam extracted from the plasma source and contributes to the increase in the thrust; these are $p_{i,t,x}$ and $p_{i,r,x}$ in Figure \ref{fig:ion_energy_flux_north_and_east_1} and $p_{e,t,x}$ and $p_{e,r,x}$ in Figure \ref{fig:electron_energy_flux_north_and_east}(A). As shown in Figure \ref{fig:ion_energy_flux_north_and_east_1}(B), the ion energy flux in the $x$-direction $p_{i,r,x}$ becomes greater as increasing the magnetic field, expected to enhance the thrust. The electron energy fluxes in the $x$-direction $p_{e,t,x}$ and $p_{e,r,x}$ are 0--65 \si{W/m^2} for the 0.4 and 2.0 kA cases as shown in Figure \ref{fig:electron_energy_flux_north_and_east}(A), indicating that the electron energy fluxes are an order of magnitude smaller than the ion energy flux. However, the electron energy flux is not negligible in the thrust generation and also increases the thruster efficiency. 

\begin{figure}
    \centering
    \includegraphics{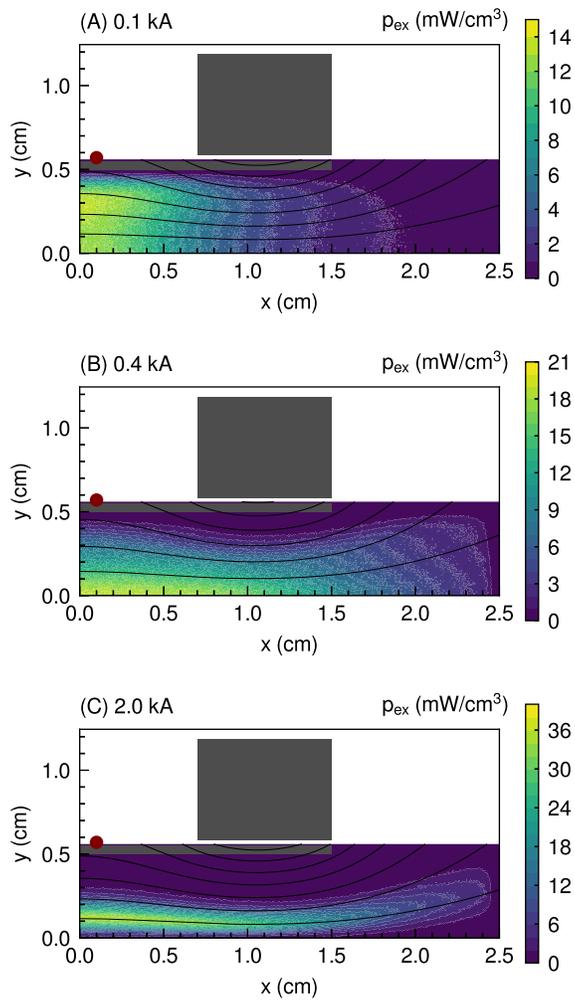}
    \caption{$x$-$y$ profiles of the excitation energy loss density $p_{ex}$ for three solenoid currents of (A) 0.1, (B) 0.4, and (C) 2.0 kA. Solid black lines show magnetic field lines.}
    \label{fig:excitation_energy_loss}
\end{figure}

\begin{figure}
    \centering
    \includegraphics{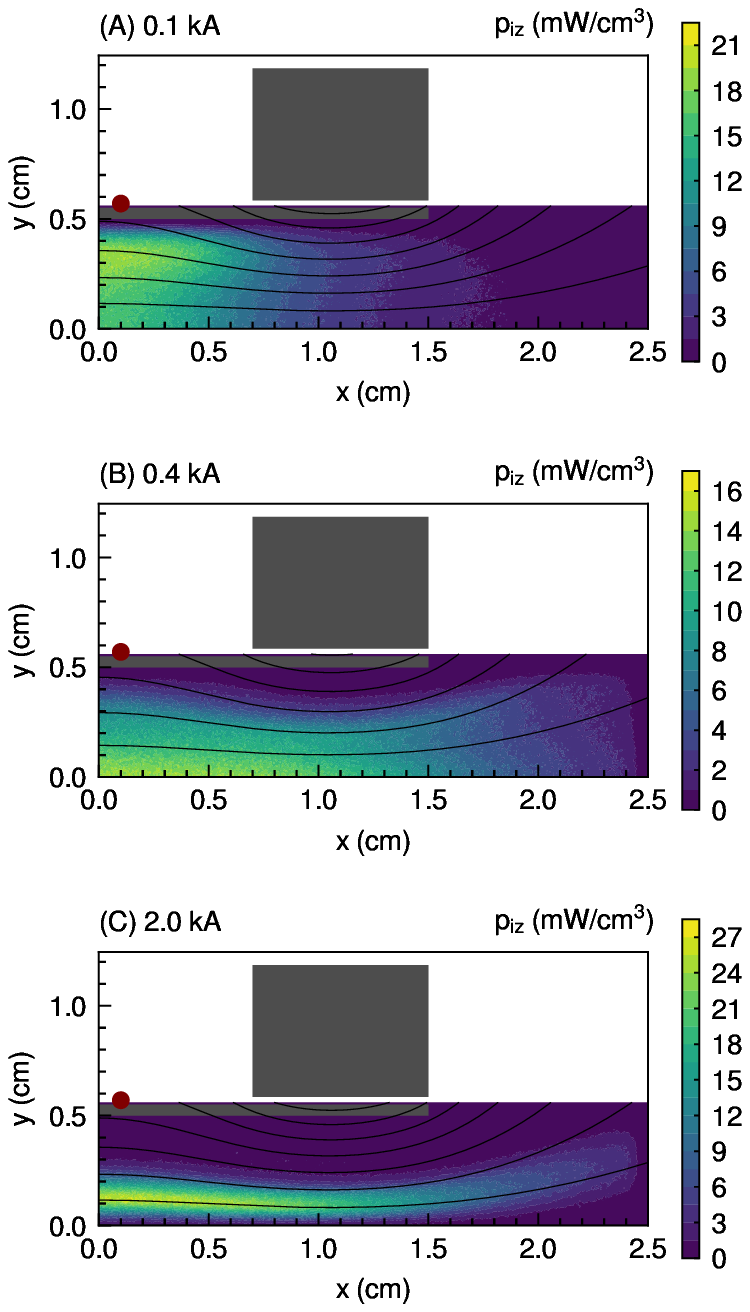}
    \caption{$x$-$y$ profiles of the ionization energy loss density $p_{iz}$ for three solenoid currents of (A) 0.1, (B) 0.4, and (C) 2.0 kA. Solid black lines show magnetic field lines.}
    \label{fig:ionization_energy_loss}
\end{figure}

Figures \ref{fig:excitation_energy_loss} and \ref{fig:ionization_energy_loss} show $x$-$y$ profiles of the excitation energy loss density $p_{ex}$ and the ionization energy loss density $p_{iz}$, respectively, for three solenoid currents of 0.1, 0.4, and 2.0 kA. $p_{ex}$ and $p_{iz}$ have similar distributions for each solenoid current and are large at the upstream region of the magnetic nozzle, where the electron number density $n_e$ is also large as shown in Figure \ref{fig:electron_number_density}.  

\begin{figure}
    \centering
    \includegraphics{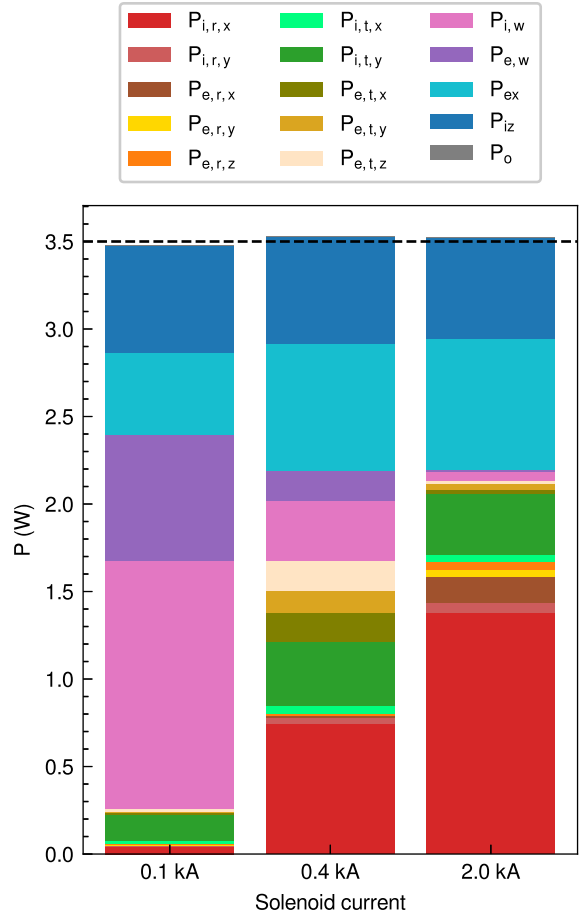}
    \caption{Energy losses in the magnetic nozzle rf plasma thruster for the three solenoid currents of 0.1 (a left bar), 0.4 (a center bar), and 2.0 kA (a right bar). The definitions of energy losses are described in Table \ref{tab:energy_definition}. $P_{o}$ contains other energy losses of $P_{i,b,z}$, $P_{i,d,z}$, and $P_{el}$, and its magnitude is less than 10$^{-3}$ W. The input power $P_{in}$ is set to 3.5 W in all solenoid currents; therefore,  the total energy losses are also approximately 3.5 W, as shown in a dashed black line.}
    \label{fig:energy_loss}
\end{figure}

Figure \ref{fig:energy_loss} shows energy losses in the magnetic nozzle rf plasma thruster for three solenoid currents of 0.1, 0.4, and 2.0 kA. The definitions of energy losses are described in Table \ref{tab:energy_definition}. $P_{o}$ contains other energy losses of $P_{i,r,z}$, $P_{i,t,z}$, and $P_{el}$, and its magnitude is less than \SI{1e-3}{W}, indicating that the elastic energy loss density and the rotation energy fluxes are negligibly small. Energy losses are calculated using Equations \eqref{eq:sum_surface_energy_loss} and \eqref{eq:sum_volume_energy_loss}. Note that the input power from the rf antenna $P_{in}$ is set to 3.5 W in all solenoid currents and the total energy losses are also almost equal to 3.5 W as a dashed black line in Figure \ref{fig:energy_loss}.

$P_{ex}$ and $P_{iz}$ are the excitation and ionization energy losses, respectively. The sums of $P_{ex}$ and $P_{iz}$ account for 31.1\%, 37.9\%, and 36.6\% for the solenoid currents of 0.1, 0.4, and 2.0 kA, respectively. Therefore, $P_{ex}$ and $P_{iz}$ account for 30--40\% regardless of the magnetic field strength. Here, $P_{ex}$ and $P_{iz}$ are considered as the minimum loss power to maintain the plasma discharge. Thus, 30--40\% of the energy absorbed by the plasma is not utilized as the plasma beam with or without the magnetic nozzle. The remaining 60--70\% of the energy can be utilized as the energy of the plasma beam, which is consistent with the maximum thruster efficiency of Hall thrusters \cite{Goebel2008_fundamentals}. Although $P_{ex}$ and $P_{iz}$ are considered unavoidable energy losses in plasma propulsion, the thruster efficiency of the magnetic nozzle rf plasma thruster can be increased to 60--70\% if the remaining energy is utilized efficiently. 

$P_{i,w}$ and $P_{e,w}$ are ion and electron energy losses by colliding with the lateral dielectric wall, respectively. The sums of $P_{i,w}$ and $P_{e,w}$ account for 61.5\%, 14.6\%, and 1.78\% for the solenoid currents of 0.1, 0.4, and 2.0 kA, respectively. Therefore, $P_{i,w}$ and $P_{e,w}$ are extremely reduced by increasing the solenoid current. This significant decrease is due to preventing the electron loss to the dielectric by the magnetic field. The sheath near the dielectric wall is also suppressed, and the ion energy loss is prevented. These results are consistent with previous experimental results reported in \cite{Takahashi2020_sr}.

$P_{i,t,x}$ and $P_{i,t,y}$ are ion energies in the $x$- and $y$-directions colliding with the top boundary, respectively, which are extracted from the plasma source but diverged by the magnetic nozzle. $P_{i,t,x}$ accounts for 0.6\%, 1.3\%, and 1.1\% for the solenoid currents of 0.1, 0.4, 2.0 kA, respectively. Therefore, $P_{i,t,x}$ is negligibly small and less contributes to the increase in the thrust regardless of the magnetic field strength. $P_{i,t,y}$ accounts for 4.31\%, 10.3\%, and 10.0\% for the solenoid currents of 0.1, 0.4, 2.0 kA, respectively. $P_{i,t,y}$ is the lowest at 0.1 kA, and it doubles at 0.4 and 2.0 kA by the magnetic nozzle. $P_{e,t,x}$, $P_{e,t,y}$, and $P_{e,t,z}$ are electron energy losses in the $x$-, $y$-, and $z$-directions colliding with the top boundary, respectively. The sum of $P_{e,t,x}$, $P_{e,t,y}$, and $P_{e,t,z}$ is relatively large for the solenoid current of 0.4 kA, while it is mostly vanished for the solenoid current of 2.0 kA, as shown in Figure \ref{fig:electron_energy_flux_north_and_east}.

$P_{i,r,x}$ and $P_{i,r,y}$ are ion energy losses in the $x$- and $y$-directions colliding with the right boundary, respectively. $P_{i,r,x}$ accounts for 1.2\%, 21.1\%, and 39.1\% for the solenoid currents of 0.1, 0.4, 2.0 kA, respectively. Therefore, $P_{i,r,x}$ dramatically increases by increasing the solenoid current, contributing to increase in the thrust. $P_{i,r,y}$ accounts for 0.1\%, 1.0\%, and 1.7\% for the solenoid currents of 0.1, 0.4, 2.0 kA, respectively. Although $P_{i,r,y}$ slightly increases with increasing the solenoid current, its effect is relatively small in the energy losses. $P_{e,r,x}$, $P_{e,r,y}$, and $P_{e,r,z}$ are electron energy losses in the $x$-, $y$-, and $z$-directions, respectively. While these electron energy losses are relatively small, they are also included in Equation \eqref{eq:thrust_energy}, somewhat contributing to the thrust generation.

\begin{figure}
    \centering
    \includegraphics{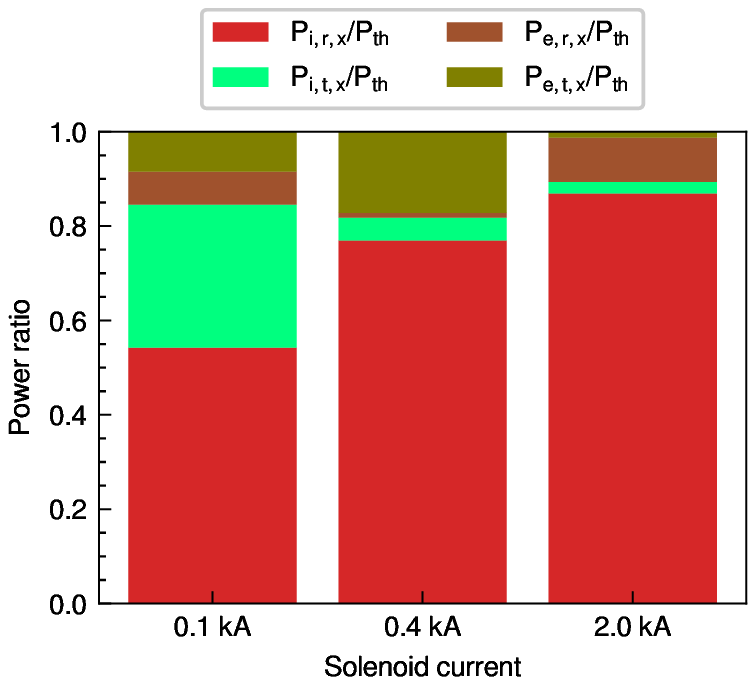}
    \caption{Power ratios to the plasma power contributing to the thrust $P_{th}$ in the magnetic nozzle rf plasma thruster for the three solenoid currents of 0.1 (a left bar), 0.4 (a center bar), and 2.0 kA (a right bar). The definitions of energy losses are described in Table \ref{tab:energy_definition}.}
    \label{fig:thrust_energy_ratio}
\end{figure}

The plasma powers contributing to the thrust $P_{th}$ are 0.07, 0.97, and 1.59 W for the solenoid currents of 0.1, 0.4, and 2.0 kA, respectively, which are calculated from Equation \eqref{eq:thrust_energy}. Then, power ratios to the plasma power contributing the thrust $P_{th}$ are calculated and shown in Figure \ref{fig:thrust_energy_ratio}. The largest component of $P_{th}$ is $P_{i,r,x}$, and the power ratios $P_{i,r,x}/P_{th}$ are 54.3\%, 77.0\%, and 86.9\% for the solenoid currents of 0.1, 0.4, and 2.0 kA, respectively. The power ratios $P_{i,t,x}/P_{th}$ are 30.3\%, 4.8\%, and 2.4\% for the solenoid currents of 0.1, 0.4, and 2.0 kA, respectively. It is implied that the divergent ion energy is suppressed by the magnetic field. The power ratios of the electron energies $(P_{e,r,x} + P_{e,t,x})/P_{th}$ are 15.5\%, 18.2\%, and 10.7\% for the solenoid currents of 0.1, 0.4, and 2.0 kA, respectively. The electron energy somewhat contributes to the thrust, while they are converted to ion energy in the sheath.

The thruster efficiencies $\eta$ for the three solenoid currents are calculated from the energy losses in the magnetic nozzle rf plasma thruster using Equation \eqref{eq:thruster_efficiency}. The thruster efficiencies are 2.1\%, 27.4\%, and 45.0\% for the solenoid currents of 0.1, 0.4, and 2.0 kA, respectively. Therefore, the thruster efficiency in the magnetic nozzle increases dramatically by increasing the solenoid current. The thruster efficiency of 2.1\% for the solenoid current of 0.1 kA is consistent with the experiment with the weak magnetic field as reported in \cite{Pottinger2011_jpd, Takahashi2011_apl, Takahashi2011_prl, Charles2012_apl, Charles2013_apl, Williams2013_jpp}. Then, the thruster efficiency of 27.4\% for the solenoid current of 0.4 kA is roughly consistent with the recent experiment in \cite{Takahashi2021_sr}, which is the maximum thruster efficiency measured to date. Although the thruster efficiency of 45.0\% for the solenoid current of 2.0 kA is not realized in experiments, the fully kinetic simulation in this paper implies that the thruster efficiency of magnetic nozzle rf plasma thrusters could achieve the thruster efficiency of 40--50\% by increasing the solenoid current. The continuous increases in the thrust has actually been observed when increasing the magnetic field up to a few kG \cite{Takahashi2016_pop}, which is much higher than that used in the thrust assessment experiment. The assessment of the thruster efficiency remains further experimental issue.

Here, the increase in the thruster efficiency with increasing the solenoid current is mainly due to the dramatic decrease of the ion and electron energy losses by colliding with the lateral dielectric wall, because the collisional energy losses and the divergent ion energy remain almost unchanged regardless of the solenoid current. The results are consistent with the previous assessment of the energy loss to the wall \cite{Takahashi2020_sr}. It is also expected that increasing the the solenoid current and preventing the energy loss by colliding with the dielectric wall improve the thruster efficiency in experiments. When the energy loss by colliding with the lateral dielectric wall is completely prevented, suppressing the divergent ions would further improves the thruster efficiency of the magnetic nozzle rf plasma thrusters.

\section*{CONCLUSION}

The fully kinetic simulations are conducted to investigate the energy losses in the magnetic nozzle rf plasma thruster for further improvement of the thruster efficiency. It is shown that the main energy loss for the weak magnetic field strength is the ion and electron energies lost to the lateral dielectric wall and they are dramatically reduced by increasing the solenoid current. The ion beam energy increases instead of the decrease of the energy loss on the wall, improving the thruster efficiency. The divergent ion energy and collisional energy losses are identified as approximately 4--12\% and 30--40\%, respectively, which are not utilized as the plasma beam. Suppressing the energy loss on the lateral dielectric wall is the key to the improvement of the thruster efficiency, and the performance of the magnetic nozzle rf plasma thrusters is expected to be further improved. 

\section*{DATA AVAILABILITY STATEMENT}

The raw data supporting the conclusions of this article will be made available by the authors, without undue reservation.

\section*{AUTHOR CONTRIBUTIONS}

KE, KT, and YT performed the conceptual design of the study. KE and YT designed the calculation model and wrote the simulation codes. KE performed the simulations. The data was analyzed by KE and discussed by KE, KT, and YT. The first draft of the manuscript was written by KE and revised by KT and YT.

\section*{FUNDING}

This work was partly supported by JSPS KAKENHI Grant Numbers JP21J15345 and JP19H00663. The computer simulation was performed on the A-KDK computer system at Research Institute for Sustainable Humanosphere, Kyoto University.

\bibliographystyle{frontiersinHLTH_FPHY}
\bibliography{main}

\end{document}